\documentclass{article}

\usepackage{mathrsfs}
\usepackage{amsfonts}
\usepackage{amsmath}
\usepackage{amssymb}
\usepackage{graphicx}

\title{The non-unique Universe}
\author{Gordon McCabe}

\begin{document}

\maketitle

\begin{abstract}

The purpose of this paper is to elucidate, by means of concepts and theorems
drawn from mathematical logic, the conditions under which the existence of a
multiverse is a logical necessity in mathematical physics, and the implications
of G\"odel's incompleteness theorem for theories of everything.

Three conclusions are obtained in the final section: (i) the theory of the
structure of our universe might be an undecidable theory, and this constitutes
a potential epistemological limit for mathematical physics, but because such a
theory must be complete, there is no \emph{ontological} barrier to the
existence of a final theory of everything; (ii) in terms of mathematical logic,
there are two different types of multiverse: classes of non-isomorphic but
elementarily equivalent models, and classes of model which are both
non-isomorphic and elementarily inequivalent; (iii) for a hypothetical theory
of everything to have only one possible model, and to thereby negate the
possible existence of a multiverse, that theory must be such that it admits
only a finite model.

\end{abstract}

\section{Introduction}

In modern mathematical physics and cosmology, a \emph{multiverse} is defined to
be a collection of possible physical universes.\footnote{This paper will
refrain from using the phrase `ensemble of universes', given that an
\emph{ensemble} is typically considered to be a space which possesses a
probability measure. It is debatable whether the universe collections
postulated by mathematical physicists and cosmologists possess a well-defined
probability measure. As Tegmark comments, ``Further work on all aspects of the
measure problem is urgently needed\ldots as this is necessary for
observationally testing any theory that involves parallel universes at any
level, including cosmological inflation and the string theory landscape,"
(2008, VIII, C).} Multiverses can be either timeless collections of disjoint,
non-interacting universes, or the result of common physical processes. The
primary examples of the latter are the universe-domains in Linde's chaotic
inflation theory (1983a and 1983b), and the universes created inside black
holes in Smolin's theory of cosmological natural selection (2006). The primary
focus of this paper, however, is on timeless multiverses.

The logical existence of such multiverses is a consequence of the fact that
mathematical physics represents the physical world by means of mathematical
structures. These are sets equipped with properties, relationships, operations
and distinguished elements, which are collectively required to satisfy certain
conditions, called the axioms of the structure. To be precise, the axioms
define a \emph{species} of structure, and each set which possesses that
structure is a member of the species. For example, the vector space axioms
define a species of structure, and each particular vector space is a member of
that species.

It is useful at the outset here to introduce some definitions from mathematical
logic, including, in particular, the distinction therein between theories and
models. A \emph{theory} $T$ is defined to be a set of sentences, in some
language, which is closed under logical implication (Enderton, p155). In other
words, any sentence which can be derived from a subset of the sentences in a
theory, is itself a sentence in the theory. A subset of sentences in a theory,
from which the entire theory can be generated by logical implication, is
referred to as a set of \emph{axioms} for the theory.\footnote{As defined in
Section 4, the subset of sentences must also be decidable.} An
\emph{intepretation} of a language provides the language with its semantics, in
the sense that it identifies: the domain over which the variables in the
language range; the elements in the domain which correspond to the constants in
the language; the elements which possess the predicates in the language; the
$n$-tuples of elements which are related by the $n$-ary relations in the
language; and the elements which result from performing $n$-ary operations upon
$n$-tuples in the domain. A \emph{model} $\mathfrak{U}$ of a theory $T$ is an
interpretation of the langauge in which that theory is expressed, which renders
each sentence in the theory as true.

In this precise sense, the axioms which define a mathematical structure are the
axioms of a theory, and each member of a species of structure is a model of
that axiomatic theory.\footnote{In this context, it should be noted that
Tegmark (1998, 2008) draws a distinction between \emph{formal systems} and
mathematical structures, rather than a distinction between theories and
models.}

If our physical universe is conceived to possess a mathematical structure, then
one can define a multiverse consisting of all the models of that species of
structure. Let us consider a couple of examples. In general relativity, a
universe is represented by a 4-dimensional differential manifold $\mathcal{M}$
equipped with a metric tensor field $g$ and a set of matter fields and gauge
force fields $\{\phi_i \}$ which generate an energy-stress-momentum tensor $T$
that satisfies the Einstein field equations
$$ T = 1/(8\pi G)(\text{Ric} - 1/2 \; \text{R} \, g) \,.
$$ $Ric$ denotes the Ricci tensor field determined by $g$,
and $\text{R}$ denotes the curvature scalar field. The matter fields have
distinctive equations of state, and include fluids, scalar fields, tensor
fields, and spinor fields. Gauge force fields, such as electromagnetism, are
described by $n$-form fields. Hence, one can define a general relativistic
multiverse to be the class of all models of such $n$-tuples
$\{\mathcal{M},g,\phi_1,...\}$, interpreted in this sense.

Alternatively, quantum field theory represents a universe to be a Lorentzian
manifold $(\mathcal{M},g)$ which is equipped with a Hilbert space
$\mathscr{H}$, a density operator $\rho$ on $\mathscr{H}$, and a collection of
operator-valued distributions $\{\hat{\phi}_i \}$ on $\mathcal{M}$ which take
their values as bounded self-adjoint operators on $\mathscr{H}$ (Wallace 2001).
A quantum field theory multiverse is the class of all models of such $n$-tuples
$\{\mathcal{M},g,\mathscr{H},\rho, \hat{\phi}_1,...\ \}$, interpreted in this
sense.

Such universe collections logically exist by virtue of the absence of
contradiction in their definition. In the style of Max Tegmark (1998, 2008),
one can then go further, and propose that these universe collections
\emph{physically} exist.

Tegmark first considers the proposal that ``some subset of all mathematical
structures\dots is endowed with\dots physical existence," (1998, p1), but
dismisses this as inadequate because it fails to explain why some particular
collection of mathematical structures is endowed with physical existence rather
than another. This is what philosophers would refer to as a problem of
\emph{contingency}, where a contingent fact is something which happens to be
true, but isn't true as a matter of necessity.

Tegmark's first multiverse paper responded to this problem of contingency by
suggesting that \emph{all} mathematical structures have physical existence.
Tegmark's 2008 paper, however, incorporated the implications of G\"odel
incompleteness and Church-Turing uncomputability, by formulating alternative
proposals that only computable structures, or finite computable structures,
physically exist (2008, p22).\footnote{Tegmark defines a computable structure
to be one whose relations can be obtained by computations which are guaranteed
to halt after a finite number of steps (2008, p20)}

In light of such speculation, the following sections analyse the concept of the
multiverse in more detail: Section 2 examines the relationship between
multiverses, theories, models and the `parameters of physics'; Section 3
considers multiverses generated by different Lagrangians; and Section 4
assesses the implications of G\"odel's incompleteness theorem for theories of
everything, and considers whether the prospect of a parameter-free theory of
everything would really negate the possible existence of a multiverse.

\section{Multiverses, parameters, theories and models}

Multiverses are often introduced by varying the so-called `parameters of
physics'. These are typically parameters in the standard model of particle
physics\footnote{Note that the standard `model' is, in terms of mathematical
logic, a theory and not a model.}, or parameters which specify the initial
conditions in general relativistic cosmology. The values of these parameters
cannot be theoretically derived, and need to be determined by experiment and
observation.

Philosopher of science Jesus Mosterin (2004) points out that ``the set of all
possible worlds is not at all defined with independence from our conceptual
schemes and models. If we keep a certain model (with its underlying theories
and mathematics) fixed, the set of the combinations of admissible values for
its free parameters gives us the set of all possible worlds (relative to that
model). It changes every time we introduce a new cosmological model (and we are
introducing them all the time). Of course, one could propose considering the
set of all possible worlds relative to all possible models formulated in all
possible languages on the basis of all possible mathematics and all possible
underlying theories, but such consideration would produce more dizziness than
enlightenment."

Mosterin's point here is aimed at the anthropic principle, and the suggestion
that there are multiverses which realise all possible combinations of values
for the parameters of physics. At face value, this might seem to be a different
type of multiverse than that obtained by varying mathematical structures and
models, but in fact, the values chosen for the free parameters of a theory
actually correspond to a choice of model.

As an example, consider the free parameters of the standard model of particle
physics. These include: the coupling constants of the strong and
electromagnetic forces; two parameters which determine the Higgs field
potential; the Weinberg angle; the masses of the elementary quarks and leptons;
and the values of four parameters in the Kobayashi-Maskawa matrix which
specifies the `mixing' of the $\{d,s,b\}$ quark flavours in weak force
interactions. In terms of a choice of model, the value chosen for the coupling
constant of a gauge field with gauge group $G$ corresponds to a choice of
metric in the Lie algebra $\mathfrak{g}$, (Derdzinksi 1992, p114-115); the
Weinberg angle corresponds to a choice of metric in the Lie algebra of the
electroweak force, (ibid., p104-111); the values chosen for the masses of the
elementary quarks and leptons correspond to the choice of a finite family of
irreducible unitary representations of the local space-time symmetry group,
from a continuous infinity of alternatives on offer (McCabe 2007); and the
choice of a specific Kobayashi-Maskawa matrix corresponds to the selection of a
specific orthogonal decomposition $\sigma_{d'} \oplus \sigma_{s'} \oplus
\sigma_{b'}$ of the fibre bundle which represents a generalization of the
$\{d,s,b\}$ quark flavours, (Derdzinski 1992, p160).

Nevertheless, Lee Smolin (2009) argues against the notion that there exists a
multiverse of (timeless) universes. Smolin believes that the need to invoke a
multiverse is rooted in the dichotomy between laws and initial conditions in
existing theoretical physics, and suggests moving beyond this paradigm.

A choice of initial conditions, however, is merely one of the means by which
particular solutions to the laws of physics are identified. More generally,
there are boundary conditions, and free parameters in the equations, which have
no special relationship to the nature of time. To reiterate, each theory in
mathematical physics represents the physical world by a species of mathematical
structure, for which there are, in general, many possible non-isomorphic
models; the laws associated with that theory select a particular sub-class of
these models. As Earman puts it, ``a practitioner of mathematical physics is
concerned with a certain mathematical structure and an associated set
$\mathfrak{M}$ of models with this structure. The\ldots laws $L$ of physics
pick out a distinguished sub-class of models $\mathfrak{M}_L := \text{Mod}(L)
\subset \mathfrak{M}$, the models satisfying the laws $L$ (or in more colorful,
if misleading, language, the models that ``obey" the laws $L$)," (p4,
2002).\footnote{If those laws contain a set of free parameters $\{p_i : i =
1,...,n\}$, then one has a different class of models $\mathfrak{M}_{L(p_{i})}$
for each set of combined values of the parameters $\{p_i\}$.} The application
of a theory to explain or predict a particular empirical phenomenon, then
requires the selection of a particular solution, i.e., a particular model. The
choice of initial conditions and boundary conditions is simply a way of picking
out a particular model of a theory.

One point of nomenclature to note in passing here is that, whilst mathematical
logicians consider a theory to be the set of sentences which define a species
of structure, physicists consider the laws which define a sub-class of
mathematical models to define a theory. If one retains the same species of
mathematical structure, but one changes the laws imposed upon it, then, as far
as physicists are concerned, one obtains a different theory. Thus, for example,
whilst general relativity represents space-time as a 4-dimensional Lorentzian
manifold, if one changes the laws imposed by general relativity upon a
Lorentzian manifold, (the Einstein field equations), then one obtains a
different physical theory.

Irrespective of the nomenclature, the crucial point is that any theory whose
domain extends to the entire universe, (i.e. any cosmological theory),
potentially has a multiverse associated with it: namely, the class of all
models of that theory. Irrespective of whether a future theory abolishes the
dichotomy between laws and initial conditions, as Smolin prescribes, the
application of that theory will require a means of identifying particular
models of the species of mathematical structure selected by the theory. If
there is only one physical universe, as Smolin claims, then the problem of
contingency will remain: why does this particular model exist and not any one
of the other possibilities? The invocation of a multiverse solves the problem
of contingency by postulating that all the possible models physically exist.

\section{Lagrangians and multiverses}

At a classical level in mathematical physics, the equations of a theory can be
economically specified by a Lagrangian, hence it is typical in physics to
identify a theory with its Lagrangian. This point is particularly crucial
because it also explains why different `Effective Field Theories' (EFTs) are
associated with different `vacua'.

The Lagrangians of particle physics typically contain scalar fields, such as
the Higgs field in the unified electroweak theory, or the moduli fields of
string theory, (which purportedly control the way in which six of the ten
space-time dimensions are compactified). The scalar fields have certain values
which constitute minima of their respective potential energy functions, and
such minima are called vacuum states (or ground states). If one assumes that in
the current universe such scalar fields reside in a vacuum state, then this can
yield new Lagrangians in two different ways. Firstly, if a particular vacuum
state is chosen and substituted into the general Lagrangian, then this yields a
reduced Lagrangian. Each different vacuum state can yield a different reduced
Lagrangian. Secondly, however, the choice of a vacuum state can yield a
Lagrangian for the low-energy fluctuations above the vacuum state. This is
called the Lagrangian of an effective field theory. As Smolin puts it, ``An
effective field theory is a semiclassical field theory which is constructed to
represent the behavior of the excitations of a vacuum state of a more
fundamental theory below some specified energy scale. They have the great
advantage that one can study a theory expanded around a particular solution,
treating that solution as a fixed background," (2005, p29). Different choices
of vacuum state will yield different effective theories of low-energy
phenomenology.

Neither the reduced Lagrangian nor an effective Lagrangian are the Lagrangian
of the fundamental theory. The selection of a vacuum state yields a new
Lagrangian, and because a Lagrangian defines a theory, the selection of a
vacuum state for a scalar field is seen to define the selection of a theory. It
is therefore typical in physics to speak, interchangeably, about the number of
possible vacua, and the number of possible effective field theories in string
theory.\footnote{Note that not all EFTs are obtained from a fundamental theory
by the selection of a vacuum state. Whilst the latter can be considered a
`top-down' approach to obtaining EFTs, there are also `bottom-up' approaches,
in which, for example, the parameters in an existing Lagrangian are modified
under Renormalization Group equations to obtain a new EFT. See Hartmann (2001)
for a comprehensive analysis of EFTs.} The collection of different string
theory vacua defines the so-called string theory `landscape', and this
landscape defines a type of multiverse.

However, it should be carefully noted that the string theory landscape defines
a collection of different (effective) theories, not a collection of models of a
fixed theory. Hence, even if one fixes a particular fundamental string theory,
and even if one selects a particular vacuum state and a particular low-energy
effective theory, this point in the string theory landscape itself corresponds
to another multiverse, consisting of the class of all models of that effective
theory.

\section{Mathematical logic, theories of everything, and multiverses}

A final theory of everything, with no free parameters, has often been
postulated as a superior alternative to the multiverse generated by our current
suite of theories, with their various free parameters. The idea here is that
the values of the free parameters in current theories, will follow by
definition from the axioms of a final theory, in the same way that the value of
pi follows from the axioms of classical Euclidean geometry. However, whilst
there may be no free parameters in a final theory, the absence of free
parameters is no guarantee that a theory will possess only one model. Hence,
even if a final, parameter-free, theory of everything is obtainable, it may
still generate a multiverse consisting of all its mutually non-isomorphic
models.

However, before we proceed to consider the conditions under which a theory of
everything will generate a multiverse, we first need to address the frequent
question of whether G\"odel's incompleteness theorem is inconsistent with the
possibility of a theory of everything.

To reiterate, theories generally have many different models. For example, each
different vector space is a model for the theory of vector spaces, and each
different group is a model for the theory of groups. The class of groups and
the class of vector spaces can be said to be species of mathematical structure.
Conversely, given any structure or model $\mathfrak{U}$, there is a theory
$\text{Th}\, \mathfrak{U}$ which consists of the sentences which are true in
the structure $\mathfrak{U}$, (Enderton, p148).

Now, a theory $T$, in the sense defined in mathematical logic, is defined to be
\emph{complete} if for any sentence $\sigma$, either $\sigma$ or its negation
$\neg \sigma$ belongs to $T$ (Enderton p156). A theory $T$ is defined to be
\emph{decidable} if there is an effective procedure of deciding whether any
given sentence $\sigma$ belongs to $T$, where an `effective procedure' is
generally defined to be a finitely-specifiable sequence of algorithmic steps,
(ibid., p61-62). A theory is \emph{axiomatizable} if there is a decidable set
of sentences in the theory, whose closure under logical implication equals the
entire theory (ibid., p156).

G\"odel's incompleteness theorem revolves around the theory of Peano arithmetic
(the theory of conventional additional and multiplicational arithmetic), and a
particular model $\mathfrak{R}=(\mathbb{N};\mathbf{0},
\mathbf{S},<,+,\cdot,\mathbf{E})$ of Peano arithmetic, whose theory
$\text{Th}\, \mathfrak{R}$ can be referred to as `number theory' (Enderton,
p182).\footnote{$\mathbb{N}$ is the set of natural numbers, $\mathbf{0}$
denotes the number zero as a distinguished element, $\mathbf{S}$ is the
successor function, $S(n) = n+1$, $<$ is the ordering relation on $\mathbb{N}$,
and $+,\cdot,\mathbf{E}$ are addition, multiplication and exponentiation.} It
transpires that the theory of Peano arithmetic is both incomplete and
undecidable. Moreover, whilst Peano arithmetic is axiomatizable, G\"odel
demonstrated that number theory $\text{Th}\, \mathfrak{R}$ is undecidable and
non-axiomatizable, (ibid., p202ff). G\"odel obtained sentences $\sigma$, which
are true in the model, but which cannot be proven from the theory of the model.
These sentences are of the self-referential form, $\sigma$ = `I am not provable
from A', where A is a subset of sentences in the theory, (ibid., p184).

It should be recognized that an incomplete theory is a highly generic
occurrence in mathematics, and is not in itself a pathology. The axiomatic
theory of groups, for example, is incomplete. Moreover, an incomplete theory
can be turned into a complete theory by adding more axioms. For example, whilst
the theory of fields is not complete, the theory of algebraically closed fields
of characteristic zero \emph{is} complete (Enderton p156). The undecidability
of a theory can also be remedied in some cases by adding more axioms, but the
crucial point is that G\"odel discovered a type of undecidability which cannot
be remedied by the addition of extra axioms.

Whilst the application of mathematics to the physical world may be fairly
untroubled by the difficulties of self-referential sentences, undecidable
sentences which are free from self-reference have been found in various
branches of mathematics. It has, for example, been established that there is no
general means of proving whether or not a pair of `triangulated' 4-dimensional
manifolds are homeomorphic (topologically identical) (Geroch and Hartle, 1986).

Any theory which includes number theory will be undecidable, hence if a final
theory of everything includes number theory, then the final theory will also be
undecidable. Given that the use of number theory is fairly pervasive in
mathematical physics, this appears to be highly damaging to the prospects for a
final theory of everything.

However, it is still conceivable that a final theory of everything might not
include number theory, and in this case, a final theory of everything could
still be both complete and decidable. In addition, even if a final theory of
everything is incomplete and undecidable, it is the models $\mathfrak{U}$ of a
theory which purport to represent physical reality, and whilst the theory of a
model, $\text{Th}\, \mathfrak{U}$, may be undecidable, it is guaranteed to be
complete. That is, every sentence in the language of the theory will either
belong or not belong to $\text{Th}\, \mathfrak{U}$.

In conclusion, the potential undecidability of $\text{Th}\, \mathfrak{U}$, the
theory of the structure of our universe, constitutes a potential
epistemological limit for mathematical physics; it is potentially a limit on
what can be proven about the structure of our universe. However, the guaranteed
completeness of $\text{Th}\, \mathfrak{U}$, entails that there is no
\emph{ontological} barrier to the existence of a final theory of everything.

The concepts of mathematical logic, introduced to explain G\"odel's theorem,
can also be exploited to shed further light on the existence of multiverses in
mathematical physics. In particular, it is fruitful to introduce the concept of
\emph{elementary equivalence}.

Two models of a theory are defined to be elementarily equivalent if they share
the same truth-values for all the sentences of the language (Enderton p97).
Whilst isomorphic models must be elementarily equivalent, there is no need for
elementarily equivalent models to be isomorphic. For example, the structure
$(\mathbb{R}, <_R)$ consisting of the real numbers, equipped with its
conventional ordering relationship, is elementarily equivalent to $(\mathbb{Q},
<_Q)$, the set of rational numbers equipped with its conventional ordering
relationship. They both provide models of a complete theory,\footnote{See
Enderton p159, for the axioms of this theory.} formulated in a first-order
language containing the symbols $=,\forall,<$. However, whilst $\mathbb{Q}$ is
a countable set, $\mathbb{R}$ is uncountable; there cannot be an isomorphic
mapping between sets of different cardinality, hence these structures are
non-isomorphic (Enderton p97-98).

Recall that any physical theory whose domain extends to the entire universe,
(i.e. any cosmological theory), potentially has a multiverse associated with
it: namely, the class of all models of that theory. Both complete and
incomplete theories are capable of generating such multiverses. Recalling that
a complete theory $T$ is one in which any sentence $\sigma$, or its negation
$\neg \sigma$, belongs to the theory $T$, it follows that every model of a
complete theory must be elementarily equivalent. A theory will in general
possess non-isomorphic models, but in the case of a complete theory its class
of non-isomorphic models will be elementarily equivalent.

In contrast, if a theory is such that there are sentences which are true in
some models but not in others, then that theory must be incomplete, and in this
case, the models of the theory will be mutually non-isomorphic \emph{and}
elementarily inequivalent.

Hence, mathematical logic suggests that the application of mathematical physics
to the universe as a whole can generate two different types of multiverse:
classes of non-isomorphic but elementarily equivalent models; and classes of
model which are both non-isomorphic and elementarily inequivalent.

The further question then arises: are there any conditions under which a theory
has only one model, up to isomorphism? In other words, are there conditions
under which a theory doesn't generate a multiverse, and the problem of
contingency (`Why this universe and not some other?') is eliminated?

The L\"owenheim-Skolem theorem, a significant result in mathematical logic
concerning the cardinalities of the models of theories, has a corollary which
provides an answer to this. This corollary holds that if a theory has a model
of any infinite cardinality, then it will have models of all infinite
cardinalities (Enderton p154). Models of different cardinality obviously cannot
be isomorphic, hence any theory, complete or incomplete, which has at least one
model of infinite cardinality, will have a multiverse associated with. (In the
case of a complete theory, the models of different cardinality will be
elementarily equivalent, even if they are non-isomorphic). Needless to say,
general relativity has models which employ the cardinality of the continuum,
hence general relativity, for example, will possess models of every
cardinality.

For a theory of mathematical physics to have only one possible model, it must
have only a finite model. A theory of everything must have a unique finite
model if the problem of contingency, and the potential existence of a
multiverse is to be eliminated.

\end{document}